\documentstyle[12pt]{article}

\def\annp #1 #2 #3 {{\em Ann.\ Phys.\ (N.Y.)} {\bf #1}, #2 (#3)}
\def\jpa #1 #2 #3 {{\em J. Phys.\ A} {\bf #1}, #2 (#3)}
\def\jsp #1 #2 #3 {{\em J. Stat.\ Phys.} {\bf #1}, #2 (#3)}
\def\pA #1 #2 #3 {{\em Physica A} {\bf #1}, #2 (#3)}
\def\prb #1 #2 #3 {{\em Phys.\ Rev.\ B} {\bf #1}, #2 (#3)}
\def\prd #1 #2 #3 {{\em Phys.\ Rev.\ D} {\bf #1}, #2 (#3)}
\def\prl #1 #2 #3 {{\em Phys.\ Rev.\ Lett.} {\bf #1}, #2 (#3)}
\begin{document}
\begin{center}
{ \Large The three-state Potts model on a triangular lattice  }

\vspace{5 mm}
Hyunggyu Park \\
Department of Physics\\
Inha University\\
Inchon, 402-751, Korea
\end{center}

\begin{abstract}
We study the phase diagram of the three-state Potts model on a triangular
lattice with
general interactions (ferro/antiferromagnetic) between nearest
neighbor spins. When the interactions along two lattice-vector directions
are antiferromagnetic and infinitely strong, this model becomes equivalent
to a six-vertex model and exhibits a first-order (KDP)
transition from an
ordered phase into a critical phase.
Comparing the excitations occurred by relaxing the restriction of
infinite-strength interactions and those in the eight-vertex model,
we analytically
obtain the critical index for those excitations and demonstrate
the existence of a critical phase for the case of finite
antiferromagnetic interactions in two directions and ferromagnetic
interactions in the other direction. When the interactions are
antiferromagnetic in all three directions, Monte Carlo simulations
show that a first-order line emerges
from the KDP point and
separates completely an ordered phase and a disordered phase.
Along the special line where all three antiferromagnetic interactions
have the same strength, the cell-spin analysis reveals that the symmetry
of the ground states is dual to the symmetry of the $n=3$ ferromagnetic
cubic model which is known to exhibit a first-order phase
transition.

\end{abstract}

\section{Introduction}

The ferromagnetic three-state Potts model has been studied extensively.
In two dimensions, its critical properties which are independent of
underlying lattices are known exactly by the
extended scaling\cite{1,2} and/or the conformal invariance\cite{3}.
When interactions between neighboring spins become antiferromagnetic,
the critical properties vary with the structure of underlying lattices.
The symmetry of the antiferromagnetic ground states is constrained by
the structure of the underlying lattice.
For example, the antiferromagnetic three-state Potts model on a square
lattice is disordered at all temperatures, but on a triangular lattice
a first-order phase transition appears at a finite temperature.
By adding  ferromagnetic next-nearest-neighbor interactions, the
antiferromagnetic three-state Potts model on a square lattice
exhibits a sequence of two
Kosterlitz-Thouless (KT) transitions\cite{4}. This model possesses a
ground-state symmetry similar to the
ferromagnetic six-state clock model\cite{5}.
In the case of mixed-type interactions, i.e.~ferromagnetic in one
direction and antiferromagnetic in the other direction,
Monte Carlo simulations\cite{6} and transfer matrix
calculations\cite{7} on a
square lattice indicate
that there is a KT-like infinite-order phase transition from a
massless low-temperature phase into a disordered phase with an essential
singularity at a finite temperature.
In this paper we examine the phase diagram of the three-state Potts
model with antiferromagnetic and mixed-type interactions on a
triangular lattice.

The three-state Potts model on a square lattice can be
mapped to a 27-vertex model on its dual lattice\cite{4} (section 2).
The triangular-lattice model with nearest-neighbor
interactions can be viewed as the square-lattice model with
nearest- and next-nearest-neighbor interactions by distorting the
lattice properly.  We map this model onto a 27-vertex model
on its dual lattice. When the interactions along two lattice-vector
directions
are antiferromagnetic and infinitely strong, only six vertex configurations
survive. From the exact solution of the six-vertex model\cite{8},
we find a first-order (KDP)
transition from an
ordered phase into a critical phase. By the stability analysis
similar to the work by den Nijs {\em et al}\cite{4}
for the square-lattice model,
we demonstrate
the existence of a critical phase for the case of finite
antiferromagnetic interactions in two directions and ferromagnetic
interactions in the other direction.
It implies that there is a KT-type transition from a critical phase
into a disordered phase in the case of mixed-type interactions.

The three-state Potts model with the isotropic antiferromagnetic
interactions has been studied previously by the real-space
renormalization\cite{9}, series expansions\cite{10},
and Monte Carlo simulations\cite{11}.
It has been shown to exhibit a strong first-order transition.
In section 3, its ground-state symmetry is investigated.
We construct the cell-spin hamiltonian by
calculating domain wall energies and show that this model can
be renormalized to the dual model of the $n=3$ cubic model\cite{12}.
The calculated values of coupling constants of this cubic model guarantee
that the phase transition of our isotropic model is of first order.

When the interactions are anisotropic,
the chirality in domain wall energies appear. In section 4, we perform
Monte Carlo simulations to understand the role of the chirality in this
model.
We find that a first-order line emerges from
the KDP point and separates completely the antiferromagnetic
ordered phase and the disordered
phase.
The chirality neither changes the nature
of the phase transition, nor gives rise to any other
phase transition in the antiferromagnetic region.
Pleliminary results for the antiferromagnetic case have been
published separately elsewhere\cite{12.1}.

We conclude in section 5 with a brief summary.

\section{Mapping to the vertex model and stability analysis}

Consider the three-state Potts model on a square lattice with
nearest-neighbor and next-nearest-neighbor interactions.
The hamiltonian of the model, $H$, is given as
\begin{equation}
\label{eq-1}
-H=\sum_{<i,j>} K \delta_{\sigma_i \sigma_j}
+ \sum_{(i,j)^\prime} L_1 \delta_{\sigma_i \sigma_j}
+ \sum_{(i,j)^{\prime\prime}} L_2 \delta_{\sigma_i \sigma_j},
\end{equation}
where $<..>$ and $(..)$ denote nearest neighbors and
next-nearest neighbors respectively (see Fig.1).
$\sigma_i$ is the Potts spin at site $i$
which takes the values of 0,1,2 and
$\delta$ is the Kronecker delta function.
The three-state Potts model on a triangular lattice with nearest-neighbor
interactions only can be obtained by taking either $L_1$ or $L_2$
to be zero. Here we take $L_2=0$.

A three-to-one mapping to a 27-vertex model on the dual lattice
is obtained by assigning arrows or zeros
to the bonds of the dual lattice. As one goes around a dual lattice site
clockwise, an outgoing (incoming) arrow is assigned
on the encountered bond if the value of the
Potts spin is increased (decreased)
by one with modulo 3
going across the bond. If the value of the Potts spin is unchanged,
we assign a zero on the encountered bond (see Fig.1).
Bolzmann weights in terms of
$u\equiv \exp(K)$ and $v\equiv \exp(L_1)$
and the number of vertices are shown in Fig.2.
In the limit $K\rightarrow -\infty$ $(u=0)$,
only six vertices are left with nonvanishing Boltzmann weights.
By normalizing Boltzmann weights with respect to the last pair
of unpolarized vertices, we find $a=1/v$ and $b=1$ where $a$ and
$b$ are the Boltzmann weights of the first two pairs of vertices.
The single parameter $\Delta$ of the six-vertex model\cite{8,13} is given
as
\begin{equation}
\label{eq-2}
\Delta=\frac{1}{2} \left( \frac{a}{b}+ \frac{b}{a} -\frac{1}{ab}
\right)=\frac{1}{2v}.
\end{equation}
{}From the exact solution of the six-vertex model\cite{8,13}, we find that
there is a first-order KDP transition at $\Delta=1$ $(v=1/2)$
from an antiferromagnetic ordered phase
$(v<1/2)$ into a critical phase $(v>1/2)$ (see Fig.3).

With finite $K$ $(u>0)$, the vertices with zeros on the bonds
can appear. Following the analysis of den Nijs {\em et al}\cite{4} for the
antiferromagnetic three-state Potts model on a square lattice,
the most important vortex excitations which drive the critical phase
into the disordered phase are the bound pairs of vortex states
(Figs.2$(e)$ and 2$(f)$) which have vorticity of $\pm 6$ (Fig.4).
The scaling dimension of these excitations can be obtained
by using the well-known relation between scaling dimensions for
excitations of different vorticities such as\cite{14}
\begin{equation}
\label{eq-3}
x_m = \left( \frac{m}{n} \right)^2 x_n,
\end{equation}
where $x_m$ is the scaling dimension for excitations of vorticity $m$
and  zero spin-wave excitation index. The scaling dimension $x_4$ is known
exactly from the Baxter's solution of the eight-vertex model\cite{15}.
Thus the scaling dimension for excitations of vorticity 6 is given
as
\begin{equation}
\label{eq-4}
x_6=\left(\frac{9}{4}\right) x_{8V},
\end{equation}
where $x_{8V}=2-y_{8V}=2-\frac{2}{\pi}\cos^{-1}(-\Delta)$.
The critical exponent $y_6=2-x_6$ becomes negative when
$v> v^*=1/(2\sin(\frac{\pi}{18}))$. So these excitations are
irrelevant in this region and the critical phase persists for
a small but finite value of $u$. For $v<v^*$, they are relevant
with respect to the $u=0$ line. In this region, these two bound pairs can
be dissociated and the system becomes disordered for any finite
value of $u$. Thus we can draw the phase diagram for small $u$ in
the axis of $v$ (see Fig.5). As $v$ increases from the
antiferromagnetically ordered phase, a first-order transition into
the disordered phase is expected near $v\simeq 0.5$
and subsequently a continuous KT-type
transition into the critical phase near $v\simeq 2.879$.
Notice that the bound pairs of vortex states in Fig.4 are
always confined for small $u$ because a string of zeros are
generated by pulling the bound pairs apart.

\section{Ground state symmetry and cell-spin hamiltonians}

First we study the ground state symmetry of the
three-state Potts model with antiferromagnetic nearest-neighbor
interactions and ferromagnetic next-nearest-neighbor interactions on
a square lattice\cite{4,5}.
Its hamiltonian is given in eq.(\ref{eq-1}) with
$K<0$ and $L=L_1=L_2>0$. This model has six equivalent ground states.
The unit cells of these ground states are shown in Fig.6.
At zero temperature, the system becomes a periodic array of
one of these unit cells. As the temperature goes higher,
the system becomes
a mixture of these unit cells and the domain walls between different
unit cells appear. The excitation energies per unit length
for these domain walls, when they are straight and do not meander,
are easy to determine
\begin{equation}
\label{eq-5}
E_{i,i+1}=L,\qquad E_{i,i+2}=2L-K/2,
\qquad\mbox{\rm and} \qquad E_{i,i+2}=2L-K,
\end{equation}
where $i$ is an integer of modulo 6 and
$E_{ij}$ is the excitation energy of the domain wall between
two ground states, $i$ and $j$. There is no chirality in domain
wall energies; $E_{ij}=E_{ji}$. We observe from eq.(\ref{eq-5}) that
the domain wall energy $E_{ij}$ depends on $|i-j|$ only
and is a periodic function of $i$ and $j$ with periodicity of 6.
This is a symmetry of the six-state
clock model where the domain wall energies depend only on the angle
between states (Fig.7).
If we assign a cell spin $\sigma$ $(\sigma=1,\cdots,6)$
for each unit cell, the cell-spin hamiltonian reduces to
the ordinary six-state ferromagnetic clock model hamiltonian
\begin{equation}
\label{eq-6}
-H=-\sum_{<i,j>} J[1-\cos\frac{2\pi}{6}(\sigma_i-\sigma_j)],
\end{equation}
with $L=-K/2=J/2$.
The six-state clock model is known to exhibit a sequence of
two KT transitions\cite{16}. This explains why there exists a critical
fan in the antiferromagnetic three-state Potts model on a square
lattice.

Now we consider the three-state Potts model
with antiferromagnetic nearest-neighbor interactions on
a triangular lattice. The coupling constants along three different
lattice-vector directions are denoted by $K_i$ $(i=1,2,3)$.
All $K_i$'s are negative.
Similar to the above square-lattice model, there are six equivalent
ground states; three up-states $U_i$ and three down-states $D_i$
$(i=1,2,3)$. Their unit cells are shown in Fig.8. We say that
the up-states have a positive helicity and the down-states a
negative helicity.
Domain wall energies
between these ground states depend on their directions and also on the
chirality.
Consider the domain wall between two up-states $U_1$ and $U_2$
in the direction 1 (Fig.9). When $U_1$ is left
and $U_2$ is right to the domain wall, the excitation energy per
unit length of
this domain wall is $-K_3$. If the ground states are interchanged,
the domain wall energy becomes $-K_2$. So when $K_2\neq K_3$, there
is a chirality in the domain wall energies.
In general, one can find
that the domain wall energy between two up-states $U_i$ and $U_j$
in the direction 1 is $E_1=-K_2\delta_{i-1,j}-K_3\delta_{i+1,j}$.
When $i=j-1$ ($j+1$), we call that the domain wall has a positive
(negative) chirality. Similarly, the domain wall energies
between two up-states in the
other directions can be obtained easily and the result is
\begin{equation}
\label{eq-7}
E_i^{\pm} (UU)=-K_{i\mp 1},
\end{equation}
where the subscript $i$ denotes the direction of the domain wall,
the superscript $\pm$ the chirality, and $UU$ in the parenthesis
represents the domain wall energy between up-states.
Repeating the same analysis on the domain walls between down-states
and also between up- and down-states, we find
\begin{eqnarray}
\label{eq-8}
E_i^{\pm}(DD) &=& -K_{i\pm 1}, \nonumber\\
E_i^{\pm}(UD) &=& -\frac{1}{3} (K_{i+1}+K_{i-1}).
\end{eqnarray}
Notice that the domain wall energies between up- and down-states
do not depend on the chirality.

The symmetry structure of the six ground states is drawn
in Fig.10. There is a ferromagnetic
chiral (or helical)
three-state Potts model symmetry\cite{17} in each triangle.
And these triangles are linked by the symmetry of a ferromagnetic
nonchiral Ising model. When $K_1=K_2=K_3$ (isotropic case),
the chirality disappears
in both triangles. Even though the number of ground states
is the same as in the square-lattice model discussed previously,
the symmetry between the ground states is completely different from
each other.

First consider the $K_1=K_2=-\infty$ $(u=0)$ limit. This is the
six-vertex model limit (see section 2). In this limit, only four
walls can survive and their energies are
\begin{equation}
\label{eq-9}
E_1^+ (UU)=E_2^- (UU)=E_1^- (DD)=E_2^+ (DD)=-K_3.
\end{equation}
Domain walls in the direction 3 are not allowed and the up-states
and down-states cannot coexist.
There will be no isolated loop excitations of domain walls
because there exist no pairs of domain walls in the same direction
with different chirality. Thus any excitation of this model can be
represented by the zig-zag lines of the domain walls of types
$E_1^{+}(UU)$ and $E_2^- (UU)$, or $E_1^- (DD)$ and $E_2^+ (DD)$
(Fig.11). The domain wall
energy is given by $v=\exp(K_3)$ and there is no energy cost at the
crossing of the walls. By mapping to the vertex model with Bolztmann
factors normalized with respect to the last pairs of the vertices,
one can recover the six-vertex model with $a=1/v$ and $b=1$ as expected
(Fig.12).
So in the $u=0$ limit, the cell-spin approach produces the exact result.

For the isotropic model $(u=v=\exp(K))$, there is no chirality.
Assign two types of cell spins, $t$ and $s$ ($t=1,2,3$ and $s=1,2$),
for each unit cell.
The $s=1$ state represents the up-states and the $s=2$ state the
down-states. Each of three states
inside the up- or down-states is represented
by the spin $t$. Then the cell-spin hamiltonian can be written as
\begin{equation}
\label{eq-10}
-H=\sum_{<i,j>}\left[ -K\delta_{s_i s_j}\delta_{t_i t_j}
+\frac{K}{3}\delta_{s_i s_j} +\frac{2}{3}K\right],
\end{equation}
which is exactly the same as the hamiltonian of the
so-called $(q_s,q_t)$ model\cite{18} with $q_s=2$ and $q_t=3$.
For $q_t=2$ it is known as the cubic model\cite{12}.
Nature of phase transitions does not depend on the underlying lattice
structure for ferromagnetic models like the above cell-spin model
($-K>0$).
So we study the above model on a square lattice which has been
investigated in details. The duality relation between the $(q_s,q_t)$
model and the $(q_t,q_s)$ model is known on a square lattice\cite{18}.
After
dropping the constant term in eq.(\ref{eq-10}), one can find the dual
hamiltonian
\begin{equation}
\label{eq-11}
-H_D=\sum_{<i,j>}\left[ D\delta_{t_i t_j}\delta_{s_i s_j}
+J\delta_{t_i t_j} \right],
\end{equation}
where
\begin{eqnarray}
\label{eq-12}
\exp (D) &=& 1+\frac{6}{\exp(-2K/3)+2\exp(K/3)-3},\nonumber\\
\exp (J) &=& 1+\frac{3[\exp(K/3)-1]}{\exp(K/3) [\exp(-K)-1]}.
\end{eqnarray}
This is the $n=3$ cubic model hamiltonian which is known to
exhibit a first-order phase transition for $J+D/2>0$ and a
continuous phase transition otherwise\cite{12}. For our model, we can prove
from eq.(\ref{eq-12}) that $D>0$ and $J+D/2>0$ for any value of
$K$. Therefore the cell-spin analysis for the antiferromagnetic
isotropic three-state Potts model on a triangular lattice shows that
there must be a first-order transition rather than a continuous transition
and the symmetry of the ground states is dual to the symmetry of
the $n=3$ cubic model. Nature of the transition is consistent
with the Monte Carlo results\cite{11}.

On the other side of the phase diagram $(K<0, L>0)$, there exists
a critical phase. There are infinitely many ground states in the
thermodynamic limit. In the ground states,
Potts spins are ordered completely along the direction 3 and
nearest-neighbor spins in the other directions should be different
(Fig.13). This model may exhibit the same critical behavior
as the square-lattice model with mixed-type
nearest-neighbor interactions only; antiferromagnetic in one direction
and ferromagnetic in the other direction\cite{6,7,19}.
The number of ground states grows exponentially
with the  linear system  size $N$; $n_G=2^N+2(-1)^N$  which is much
smaller
than the square-lattice
model with antiferromagnetic interactions in both directions.
So this model may not be disordered at all temperatures in contrast to
the square-lattice anitiferromagnetic model.
Our stability analysis in the previous section suggests that the
transition into the critical phase is of KT type, which is consistent with
numerical results by Monte Carlo simulations\cite{6} and transfer matrix
calculations\cite{7} for the square-lattice model with mixed-type
interactions.

\section{Monte Carlo simulations and phase diagram}

Consider the antiferromagnetic region of the phase diagram
$(0<u,v<1)$.
Along the $u=v$ line (isotropic case), it is shown in the previous
section that the antiferromagnetic three-state Potts model on a
triangular lattice can be renormalized, by the cell-spin approximation,
to the dual model of the $n=3$ cubic model and then should exhibit
a first-order phase transition.

For the anisotropic model, the chirality
between ground states appear. There has been some interests in the role
of the chirality in the ordinary ferromagnetic Potts models\cite{17,20}.
Even though
the chiral operator is relevant for the three-state Potts model, it does
not introduce a new independent exponent, i.e.~$x_{CH}=x_T+1$ where
$x_{CH}$ and $x_T$ are the chiral and temperature scaling dimension
respectively\cite{21}. So it is believed that the scaling behavior does not
change in the presence of the chirality for the three-state Potts model.
This has been shown analytically for the hard hexagon model\cite{22} and
numerically for the chiral three-state Potts model\cite{17}
and the triangular Ising lattice gas\cite{23}.
The ground-state structure of our anisotropic model
is much more complicated than that of the chiral
three-state models. As explained in the previous section,
it has two chiral three-state model symmetries linked by the
Ising symmetry. So it may be quite interesting to find out
what kind of role the chirality plays in this model. We use Monte
Carlo simulations to investigate the phase diagram of this model.

We run conventional heat-bath Monte Carlo simulations on a
$60\times 60$ triangular lattice along the $u=v$, $u^2=v$, $u=v^2$, and
$u^3=v$ lines. Typically a few $10^4$ Monte Carlo steps per spin (MCS)
are performed for a given value of $u$ and $v$.
We measure the antiferromagnetic order parameter $m_{AF}$\cite{11}
and the energy
density $e$. The order parameter $m_{AF}$ is defined as
\begin{equation}
\label{eq-13}
m_{AF}=\left\langle \frac{3}{2} \sum_{\alpha\neq\beta\neq\gamma}
\left( \frac{N_\alpha^A+N_\beta^B+N_\gamma^C}{N_t}-\frac{1}{3}\right)^2
\right\rangle^{\frac{1}{2}},
\end{equation}
where $N_\alpha^X$ is the number of spins of state $\alpha$ in the
sublattice $X$ and $N_t$ is the total number of spins.
Unfortunately,
along all four lines, we find a very strong first-order phase transition
from an antiferromagnetically ordered phase into a disordered phase.
In Table I, we list the values of the coupling constant $u$ at the
first-order transitions
and the jump of  the order parameter and  the energy density.
The five first-order transition points (including the KDP point)
can be connected by a smooth line which
starts from the KDP point $(u,v)=(0,\frac{1}{2})$ and apparently ends
at the point $(0.28,0)$ (Fig.14).
As the anisotropy increases,
the magnitude of the order-parameter jump becomes bigger.
Our results imply
that the strong first-order transition due to the cubic
nature of this model
preempts all possible continuous transitions in the whole
antiferromagnetic region.

We also run Monte Carlo simulations in the region of mixed-type
interactions where a critical phase is expected.  The specific heat is
measured along the $u=1/v$ line. We find a characteristic of the
KT transition in the shape of the specific heat, which has a very
broad bump well outside of the expected critical phase.
This is consistent with
our analytical result which suggests the existence of
KT transitions in section 2 and also with some numerical results
for the square-lattice model with mixed-type interactions\cite{6,7,19}.

\section{Summary}

We investigate
the phase diagram of the three-state Potts model on a triangular lattice
with ferro/antiferromagnetic nearest neighbor interactions. In the limit
of infinite antiferromagnetic interactions along two lattice-vector
directions, this model maps onto the six-vertex model and we find the KDP
first-order transition from an antiferromagnetic ordered phase into
a critical phase. By the stability analysis, we demonstrate the existence
of the KT transition from a critical phase into a disordered phase for
the case of finite antiferromagnetic interactions in two directions and
ferromagnetic interactions in the other direction. The critical parameter
in the limit of infinite antiferromagnetic interactions is analytically
obtained.
In this case the ground
states are equivalent to those of the three-state Potts model on a square
lattice with mixed-type interactions. Our result implies that this
square-lattice model should have the same kind of the KT transition, which
confirms previous numerical results\cite{19}.

When the interactions are antiferromagnetic in all three directions, we
find six ground states. These ground states have
two chiral three-state model
symmetries linked by the Ising symmetry. For the isotropic model, the
chirality disappears and the cell-spin analysis reveals that the
isotropic model can be renormalized to the dual model of the $n=3$
ferromagnetic cubic model with coupling constants guaranteeing the
first-order phase transition. Monte Carlo simulations for
the anisotropic model shows that a first-order line emerges from the
KDP point and separates completely the antiferromagnetic ordered phase
and the disordered phase, which indicates that
the chirality is not relevant in this model.

\vspace{5mm}
\noindent{\bf Acknowledgements}
\vspace{3mm}

This work was initiated when the author was at the
University of Washington.
He thanks M.~den Nijs and T.~T.~Truong for useful discussions
and T. C. Chey for his kind support.
This work is supported in part by
the Korean Science and Engineering Foundation
through the SRC program of SNU-CTP and also through
the Center for Thermal
and Statistical Physics at Korea University.

\newpage
\begin{center}
{\Large {\bf Table Caption}}
\vspace{5mm}
\end{center}
\begin{description}
\item[{\bf Table 1 :}]
Numerical values of
the coupling constant $u_t$, the order-parameter jump $\Delta m_{AF}$,
and the energy-density jump $\Delta e$ at the first-order transitions.
Numbers in parentheses represent the errors in the last digits.
\end{description}
\begin{center}
\begin{tabular}{|c||c|c|c|c|}
\hline
             & $u=v$ & $u^2=v$ & $u=v^2$ & $u^3=v$ \\ \hline\hline
$u_t$        & $0.205(1)$& $0.263(2)$& $0.120(2)$& $0.278(3)$   \\ \hline
$\Delta m_{AF}$& $0.70(3)$ & $0.81(3)$ & $0.78(2)$ & $0.88(2)$   \\ \hline
$\Delta e$     & $0.17(2)$ & $0.20(2)$ & $0.21(1)$ & $0.24(1)$   \\ \hline
\end{tabular}
\end{center}

\newpage
\begin{center}
{\Large {\bf Figure Captions}}
\vspace{5mm}
\end{center}
\begin{description}
\item[{\bf Fig.1 :}]
A typical configuration of Potts spin on a square lattice (thick line)
and its corresponding vertex configuration on its dual lattice (thin
line). Coupling constants of nearest-neighbor ($K$) and
next-nearest-neighbor interactions ($L_1$, $L_2$) between Potts
spins are shown.
\item[{\bf Fig.2 :}]
The 27-vertex respresentation on a square lattice for the three-state
Potts model on a triangular lattice: vertex types, their Boltzmann
factors, and the number of each vertex type.
\item[{\bf Fig.3 :}]
Phase diagram of the six-vertex model. The dashed line corresponds
to our model at $u=0$.
\item[{\bf Fig.4 :}]
Bound pairs of vortex states in Figs.2($e$) and 2($f$).
\item[{\bf Fig.5 :}]
Phase diagram of our model for small $u$.
\item[{\bf Fig.6 :}]
Unit cells of the six ground states of the square-lattice model.
\item[{\bf Fig.7 :}]
Ground-state symmetry of the ferromagnetic six-state clock model.
\item[{\bf Fig.8 :}]
Unit cells of the six ground states of the triangular-lattice model:
$U_i$ (up-state) has a positive helicity and $D_i$ (down-state) a
negative helicity.
\item[{\bf Fig.9 :}]
The domain wall between $U_i$ and $U_j$ ground states in the direction 1.
Here $i^\prime=i+1$, $i^{\prime\prime}=i-1$,
$j^\prime=j+1$, $j^{\prime\prime}=j-1$.
\item[{\bf Fig.10 :}]
Ground-state symmetry of the triangular-lattice model.
\item[{\bf Fig.11 :}]
Domain walls (dashed lines) on a distorted triangular lattice at
$u=0$. The thick bonds are unhappy (ferromagnetic) bonds which
domain walls go across. Notice that
there are no unhappy bonds where two domain walls cross.
\item[{\bf Fig.12 :}]
Six vertex types of domain walls (dashed lines) with unnormalized
Boltzmann factors. Solid lines represent no domain walls.
\item[{\bf Fig.13 :}]
One of the infinitely many ground states for the model with mixed-type
interactions.
\item[{\bf Fig.14 :}]
Phase diagram of the antiferromagnetic three-state Potts model on a
triangular lattice.
\end{description}

\end{document}